\documentclass[twocolumn,english,aps,twocolumn]{revtex4}
\usepackage[T1]{fontenc}
\usepackage[latin9]{inputenc}
\usepackage{textcomp}
\usepackage{amsthm}
\usepackage{amsmath}
\usepackage{graphicx}
\usepackage{amssymb}
\usepackage{esint}

\makeatletter
\@ifundefined{textcolor}{}
{%
 \definecolor{BLACK}{gray}{0}
 \definecolor{WHITE}{gray}{1}
 \definecolor{RED}{rgb}{1,0,0}
 \definecolor{GREEN}{rgb}{0,1,0}
 \definecolor{BLUE}{rgb}{0,0,1}
 \definecolor{CYAN}{cmyk}{1,0,0,0}
 \definecolor{MAGENTA}{cmyk}{0,1,0,0}
 \definecolor{YELLOW}{cmyk}{0,0,1,0}
 }

\makeatother

\usepackage{babel}

\begin{document}

\title{Storage and Retrieval of a Microwave Field in a Spin Ensemble}

\author{Y. Kubo$^{1}$, I. Diniz$^{2}$, A. Dewes$^{1}$, V. Jacques$^{3}$,
A. Dréau$^{3}$, J.-F. Roch$^{3}$, A. Auffeves$^{2}$, D. Vion$^{1}$,
D. Esteve$^{1}$, and P. Bertet$^{1}$}

\affiliation{$^{1}$Quantronics group, SPEC (CNRS URA 2464), IRAMIS, DSM, CEA-Saclay,
91191 Gif-sur-Yvette, France }

\affiliation{$^{2}$Institut Néel, CNRS, BP 166, 38042 Grenoble, France }

\affiliation{$^{3}$LPQM (CNRS UMR 8537), ENS de Cachan, 94235 Cachan, France}

\date{\today}
\begin{abstract}
We report the storage and retrieval of a small microwave field from
a superconducting resonator into collective excitations of a spin
ensemble. The spins are nitrogen-vacancy centers in a diamond crystal.
The storage time of the order of $30$~ns is limited by inhomogeneous
broadening of the spin ensemble. 
\end{abstract}
\maketitle
Superconducting qubits are promising candidates for quantum information
processing; however their coherence times \cite{Paik2011} cannot
yet compete with those of microscopic systems such as atoms \cite{WinelandRMP},
electron or nuclear spins \cite{NVLongCoherence}. Hybrid quantum
circuit architectures have thus been proposed \cite{HybridsMolmer,HybridsKurizki,HybridsZoller,HybridsImamoglu,HybridsSchmiedmayer},
in which microscopic systems would be used as quantum memory for superconducting
qubits. Whereas the coupling of one individual atom or spin to a superconducting
circuit is usually too weak, the coupling constant of an ensemble
of $N$ such systems is enhanced by $\sqrt{N}$, allowing to reach
the strong coupling regime requested for quantum information applications.
Proposals for spin-ensemble based hybrid quantum circuits often consist
of a superconducting resonator used as a quantum bus between the ensemble
and the superconducting qubit. On the experimental side \cite{Kubo2010,Schuster2010,Majer2011,Bushev2011},
strong coupling between a spin ensemble and a superconducting resonator
has up to now been demonstrated only spectroscopically. Here we report
the first time-domain measurements of the coherent storage and retrieval
of a classical microwave field \cite{BriggsEnsemble} of about $100$
photons from a superconducting resonator into collective excitations
of a spin ensemble consisting of negatively-charged nitrogen-vacancy
centers in diamond (NV centers), an important step towards a spin-based
hybrid quantum circuit architecture. 

The experiment relies on the fact that the interaction between the
electromagnetic field in the cavity mode and the spin ensemble involves
only one collective spin variable, which behaves as a harmonic oscillator
in the limit of low excitation energy \cite{HybridsMolmer}. This
effective spin oscillator is magnetically coupled with a collective
coupling constant $g_{\mathrm{ens}}$ to the superconducting resonator
whose frequency can be tuned. When two such coupled harmonic oscillators
are suddenly put into resonance, they coherently exchange energy with
a period $\pi/g_{\mathrm{ens}}$. We observe this dynamics by measuring
the amplitude of the microwave field leaking out of the resonator
after its interaction with the spins, which is found to oscillate
as a function of the interaction time. This storage-retrieval cycle
is however damped in a relatively short time, which as we show is
limited by the inhomogeneous broadening of the NV center ensemble.
Quantitative agreement with recent theoretical work \cite{Diniz2011,Kurucz2011}
is obtained for a consistent set of data, covering spectroscopic as
well as time-domain measurements.

\begin{figure}
\includegraphics[width=8.3cm]{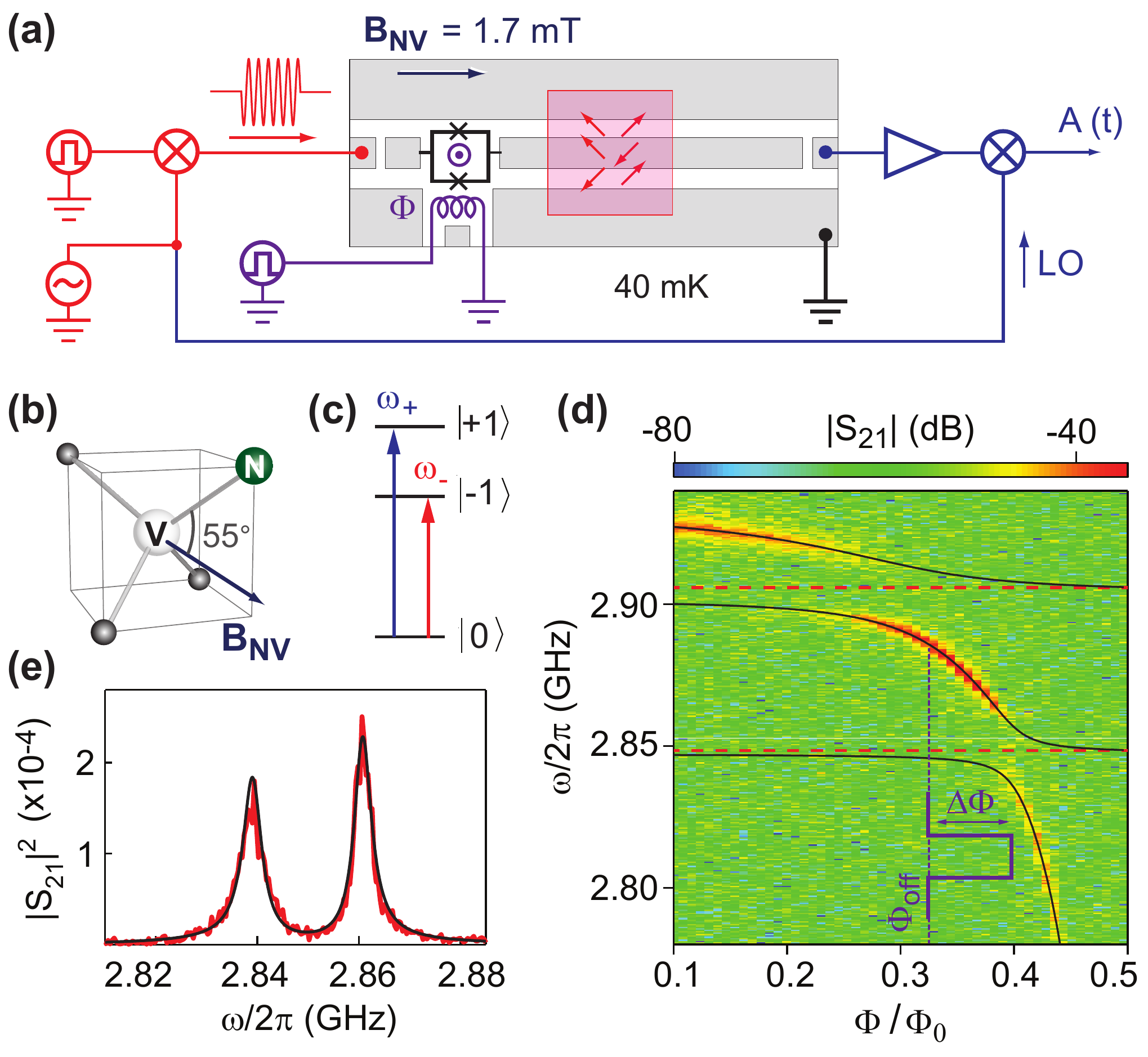}\caption{(a) Setup description. The NV centers ensemble is magnetically coupled
to the coplanar resonator containing a SQUID. The flux $\Phi$ through
the SQUID loop can be tuned on a nanosecond scale by applying current
pulses to an on-chip antenna. A magnetic field $B_{\mathrm{NV}}=1.7$~mT
is applied to the spins, parallel to the sample and to the $[1,0,0]$
crystal axis (b). The amplitude $A(t)$ of microwave pulses transmitted
through the resonator is measured by homodyne detection at room-temperature
after amplification. (c) Simplified energy level scheme of a NV center.
(d) Resonator transmission $\left|S_{21}\right|(\omega)$ at $B_{\mathrm{NV}}=1.7$~mT
as a function of $\Phi$ (in units of the superconducting flux quantum
$\Phi_{0}=h/2e$) showing two vacuum Rabi splittings. The solid line
is a fit to the data using the coupled oscillators model \cite{Kubo2010}.
(e) Vacuum Rabi splitting close to resonance with $\omega_{-}$. Black
line is experimental data, red line is theory as explained in the
text (rescaled in amplitude to fit the data), assuming a spins-resonator
detuning of $0.5$~MHz.}

\label{Fig1}%
\end{figure}

In our experiment \cite{Kubo2010} (see Fig.\ref{Fig1}a and Supplementary
Material), an ensemble of $\sim10^{12}$~NV centers in a diamond
crystal are magnetically coupled to a superconducting resonator. A
static magnetic field $B_{\mathrm{NV}}=1.7$~mT is applied to the
spins, parallel to the chip surface, along the $[1,0,0]$ crystal
axis within a few degrees (see Fig.\ref{Fig1}b). With this orientation,
the four possible NV center crystalline orientations all make approximately
the same angle $\theta\simeq55\text{\textdegree}$ with $B_{\mathrm{NV}}$
so that their resonance frequencies (see Fig.\ref{Fig1}c) are approximately
equal. The resonator includes a four-SQUID array so that its frequency
$\omega_{\mathrm{r}}(\Phi)$ can be tuned by changing the flux $\Phi$
through the SQUIDs loop \cite{TunableResonatorsPalacios}. Such a
frequency tuning can be done on a nanosecond timescale \cite{Sandberg}.
For time-domain experiments, we measure the amplitude $A(t)$ of microwave
pulses transmitted through the resonator using homodyne detection
followed by sampling and averaging.

We first characterize the resonator-spins system by measuring the
resonator transmission while scanning $\Phi$. As shown in Fig.\ref{Fig1}d,
two vacuum Rabi splittings are observed when the resonator frequency
matches either of the two NV center transitions $\omega_{-}$(resp.
$\omega_{+}$) from ground state $m_{\mathrm{S}}=0$ to $m_{\mathrm{S}}=-1$
(resp. $m_{\mathrm{S}}=+1$). A fit of these data using a coupled
oscillators model \cite{Kubo2010} yields $g_{\mathrm{ens}}/2\pi=10.6$~MHz
for the lower frequency anticrossing occurring at $\omega_{-}/2\pi=2.85$~GHz,
on which we will focus in the following. The transmission close to
the middle of the anticrossing is shown in Fig. \ref{Fig1}e where
two well-resolved polaritonic peaks can be seen, an indication that
the two oscillators are in the strong coupling regime and that the
coherent exchange of excitations between the resonator and the spin
ensemble can be observed in the time-domain. 

\begin{figure}
\includegraphics[width=8.35cm]{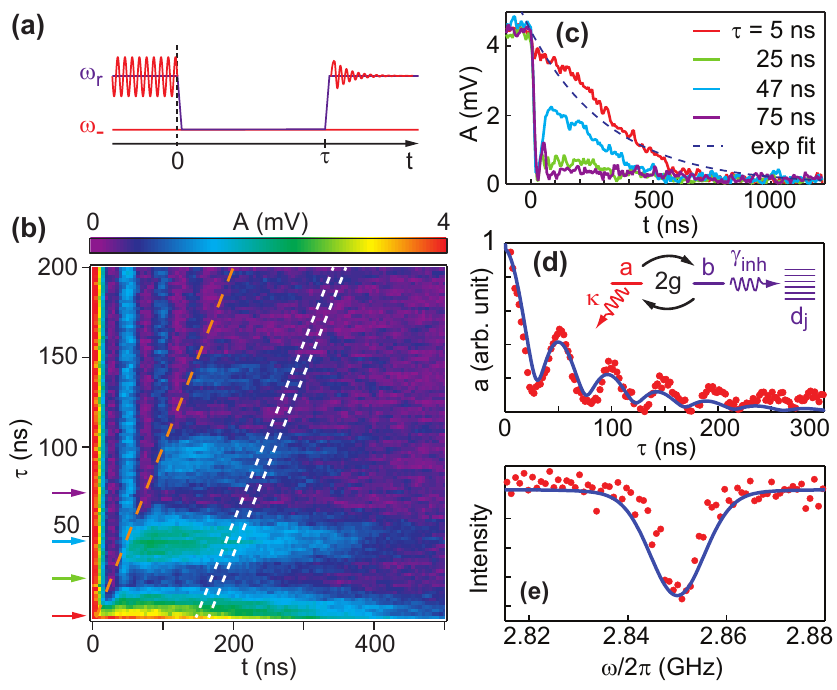}\caption{(a) Sketch of an experimental sequence (b) Output amplitude $A(t)$
measured for increasing flux pulse duration $\tau$, for a flux pulse
amplitude such that $\omega_{r}(\Phi_{\mathrm{OFF}}+\overline{\Delta\Phi})=\omega_{-}$.
The dashed orange line indicates $t=\tau$. The two white dashed lines
indicate the time window on which averaging is performed to compute
$a(\tau)$ (see text). (c) Amplitude $A(t)$ for $g_{\mathrm{ens}}\tau=0,\pi,2\pi,3\pi$.
(d) Normalized amplitude $a(\tau)=\overline{A(\tau+t_{\mathrm{off}})}$
(see text). Red dots are experimental data, continuous blue line is
theory as explained in the text. (inset) Sketch of the physical process
leading to damping of the oscillations because of coupling of the
superradiant state to the bath of subradiant states. (e) ODMR signal
measured at room-temperature (red dots), compared to the spin distribution
$\rho(\omega)$ (see text) used in the theoretical calculations (blue
line). }

\label{Fig2}%
\end{figure}

To demonstrate such dynamics, the experiment proceeds as follows (see
Fig.\ref{Fig2}a). At the beginning of each experimental sequence
(averaged out typically $5\cdot10^{5}$ times), the resonator is biased
at a flux $\Phi_{\mathrm{OFF}}$ such that its frequency $\omega_{\mathrm{r}}(\Phi_{\mathrm{OFF}})/2\pi=2.88$~GHz
is out of resonance with the spins. A microwave pulse at frequency
$\omega_{\mathrm{r}}(\Phi_{\mathrm{OFF}})$ of duration $2\mu$s (much
longer than the resonator damping time $T_{\mathrm{cav}}=Q/\omega_{\mathrm{r}}\sim100$~ns)
establishes a steady-state coherent field of amplitude $\alpha$ inside
the resonator (with $\left|\alpha\right|^{2}\sim120$ photons). Right
after the microwave pulse is switched off ($t=0$), the resonator
frequency is brought close to $\omega_{-}$ by a flux pulse of amplitude
$\Delta\Phi$ (see Supplementary Material) and duration $\tau$, during
which the resonator and the spin ensemble may exchange energy causing
the intraresonator field amplitude $|\alpha(t)|$ to oscillate. After
the flux pulse, the only evolution of the field is an exponential
decay $|\alpha(t>\tau)|=|\alpha(\tau)|\exp(-(t-\tau)/2T_{\mathbf{\mathrm{cav}}})$.
Measuring the amplitude of the exponentially damped microwave signal
that leaks out of the cavity therefore directly yields $|\alpha(\tau)|$
and reveals the coupled resonator-spins dynamics. 

Figure \ref{Fig2}b shows the results obtained for a flux pulse amplitude
$\overline{\Delta\Phi}$ that puts the resonator in resonance with
the spins at $\omega_{-}$. The two-dimensional plot shows the measured
microwave output amplitude $A(t)$ for increasing $\tau$. The curve
$A(t)$ for the shortest flux pulse $\tau=5$~ns (see Fig. \ref{Fig2}c)
showing the microwave field decay after the pulse is switched off
is only approximately exponential, due to a slight nonlinearity of
the resonator caused by the presence of the SQUID \cite{Ong2011},
which we neglect in the following. For increasing $\tau$, the curves
$A(t)$ clearly display two parts : from $t=0$ to $t=\tau$ rapid
transient oscillations are observed, which are unfortunately difficult
to quantitatively interpret due to the finite bandwidth of our homodyne
detection setup. After $t=\tau$, $A(t)$ shows a decay similar to
the one observed for the shortest pulse but with an amplitude that
oscillates with $\tau$ with a period $T\sim50$~ns, in quantitative
agreement with the coupling strength estimated from the vacuum Rabi
splitting $T\simeq\pi/g_{\mathrm{ens}}$. This establishes that the
observed oscillation is indeed due to cycles where the microwave field
is first stored into a collective spin excitation ($0<g_{\mathrm{ens}}\tau<\pi$)
then retrieved ($\pi<g_{\mathrm{ens}}\tau<2\pi$). Curves corresponding
to various steps of the cycle are also shown in Fig. \ref{Fig2}c.
Note that by using a homodyne detection scheme we can only measure
a field that has a well-defined phase relation with the local oscillator,
implying that phase coherence is indeed preserved even after several
(storage, retrieval) cycles. A full quantum state tomography would
however be needed in order to assess the fidelity of the field storage
at the quantum level, which is beyond the scope of the present work. 

In order to obtain more quantitative insight into these data, we compute
for each $\tau$ the quantity $a(\tau)=\overline{A(\tau+t_{\mathrm{off}})}$
where $t_{\mathrm{off}}=140$~ns is an offset that was chosen to
avoid taking into account transients and where an additional time-averaging
is performed on a $20$~ns window (see Fig. \ref{Fig2}b). Since
$a(\tau)$ is directly proportional to $|\alpha(\tau)|$ it is a good
quantity to investigate the spin-resonator dynamics. It is plotted
in Fig.\ref{Fig2}d for the previous flux pulse parameters. Several
cycles are observed before the oscillation amplitude and offset decay
simultaneously to zero. The fact that the field amplitude decays to
zero within a hundred of nanoseconds might come as a surprise given
the fact that the energy damping time of individual spins is in the
millisecond range at room-temperature and is expected to be well above
one second at low temperature \cite{Majer2011}. The only energy dissipation
should thus be caused by damping of the resonator, which occurs on
a significantly longer timescale. 

Understanding this decay requires to take into account the inhomogeneous
broadening of the spin ensemble. Following \cite{Kurucz2011,Diniz2011},
we model each spin in the Holstein-Primakoff approximation (valid
in our experiment which is deep in the low-excitation regime) by a
harmonic oscillator of frequency $\omega_{\mathrm{j}}$ with annihilation
(creation) operator $b_{\mathrm{j}}$ ($b_{\mathrm{j}}^{\dagger}$),
the resonator field being described by the annihilation (creation)
operator $a$ ($a^{\dagger}$). The system Hamiltonian is then $H/\hbar=\omega_{\mathrm{r}}(\Phi)a^{\dagger}a+\sum\omega_{\mathrm{j}}b_{\mathrm{j}}^{\dagger}b_{\mathrm{j}}+\sum g_{\mathrm{j}}(b_{\mathrm{j}}^{\dagger}a+b_{\mathrm{j}}a^{\dagger})$,
$g_{\mathrm{j}}$ being the coupling constant of spin $j$ with the
resonator. Turning to a new orthogonal basis consisting of the \emph{superradiant}
spin-wave mode $b=(1/g_{\mathrm{ens}})\sum g_{\mathrm{j}}b_{\mathrm{j}}$
of frequency $\omega_{-}$ and $N-1$ orthogonal \textit{dark} modes
$d_{\mathrm{j}}$ of frequency $\tilde{\omega}_{\mathrm{j}}$, with
$g_{\mathrm{ens}}=(\sum\left|g_{\mathrm{j}}\right|^{2})^{1/2}$, the
Hamiltonian becomes $H/\hbar=\omega_{\mathrm{r}}(\Phi)a^{\dagger}a+\omega_{\mathrm{-}}b^{\dagger}b+g_{\mathrm{ens}}(a^{\dagger}b+h.c)+\sum\tilde{\omega}_{\mathrm{j}}d_{\mathrm{j}}^{\dagger}d_{\mathrm{j}}+\sum\gamma_{\mathrm{j}}(b^{\dagger}d_{\mathrm{j}}+h.c)$
with $\gamma_{\mathrm{j}}\sim\Delta/\sqrt{N}$. Here $\Delta$ is
the variance of the $\omega_{j}$ distribution, i.e. the inhomogeneous
spins linewidth. Note that in this expression we have neglected terms
coupling the dark modes together. From the previous expression we
see that the superradiant spin-wave mode is the only one coupled to
the resonator field (with constant $g_{\mathrm{ens}}$), but that
in the presence of inhomogeneous broadening ($\Delta>0$) it is also
coupled to the $N-1$ dark modes. In the large-$N$ limit, these dark
modes act as a bath into which energy is damped at a rate $\gamma_{\mathrm{inh}}$,
explaining the fast decay in Fig.\ref{Fig2}d. Introducing the normalized
spin density distribution $\rho(\omega)$ (centered around $0$ instead
of $\omega_{-}$), one can show that $\gamma_{\mathrm{inh}}\simeq2\pi\rho(g_{\mathrm{ens}})g_{\mathrm{ens}}^{2}$
\cite{Diniz2011} in the $g_{\mathrm{ens}}\gg\Delta$ limit. An important
consequence of this formula is that increasing $g_{\mathrm{ens}}$
(by e.g. increasing the spins density) actually helps reducing $\gamma_{\mathrm{inh}}$
provided $\rho(\omega)$ decays faster than $\omega^{-2}$ \cite{Diniz2011,Kurucz2011}.
This implies that a strong coupling to the cavity protects the system
against the consequences of inhomogeneous broadening, an effect called
\textit{cavity protection} in \cite{Diniz2011}. In our experiment
however we are not in the $g_{\mathrm{ens}}\gg\Delta$ limit (see
below) and these approximate formula are not valid. We thus resort
to explicit analytical formula obtained using input-output relations
on $H$ \cite{Diniz2011}, which allow us to calculate both the resonator
transmission $t(\omega)=\kappa/[2i(\omega-\omega_{\mathrm{r}})-\kappa-2iW(\omega)]$
with $W(\omega)=g_{\mathrm{ens}}^{2}\int_{-\infty}^{+\infty}\rho(\omega')d\omega'/[\omega-\omega']$
and the time-domain signal as the modulus of the inverse Fourier-Laplace
transform of $t(\omega)$ (see \cite{Diniz2011} and Supplementary
Material). 

\begin{figure}
\includegraphics[width=7cm]{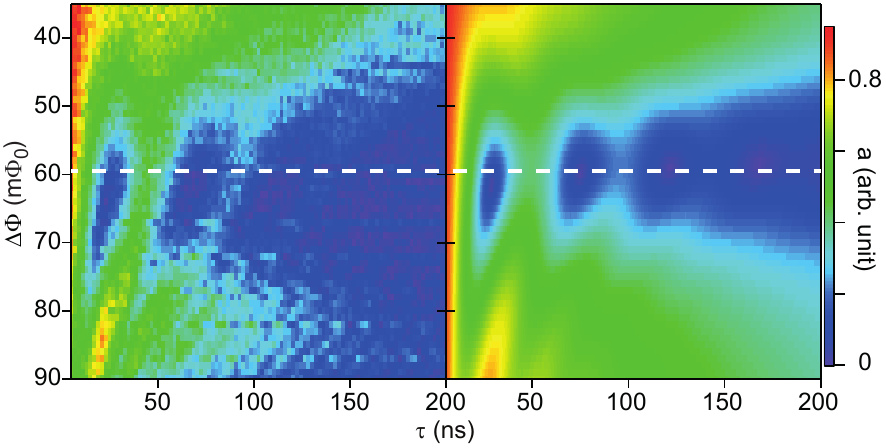}\caption{Coherent field exchange between the resonator and the spins for varying
flux pulse amplitude $\Delta\Phi$. Normalized amplitude $a(\tau)$
(see text) is plotted versus $\tau$ and $\Delta\Phi$. Left graph
is experimental data, right graph is theory. The white dashed line
indicates $\overline{\Delta\Phi}$.}

\label{Fig3}%
\end{figure}

To estimate these quantities in our experiment, we need to know the
spin density distribution $\rho(\omega)$. In NV centers ensembles,
the inhomogeneous linewidth is caused by dipolar interactions with
neighboring spins, either paramagnetic nitrogen impurities (so-called
P1 centers) that were not converted into NV centers during the sample
processing, or $^{13}C$ nuclei that are present in natural abundance
($1.1$\%) \cite{CoherenceTime}. Note that neighboring NV centers
do not contribute to the inhomogeneous linewidth because at the temperature
of our experiment they are frozen in the $m_{\mathrm{S}}=0$ state.
In our sample, a FWHM linewidth $\Delta/2\pi\sim7$~MHz was measured
by optically-detected magnetic resonance (ODMR) at room-temperature
(see Fig.\ref{Fig2}e), compatible with the linewidth expected from
its nominal P1 centers concentration $\sim100$~ppm \cite{CoherenceTime}.
As shown in \cite{Kurucz2011,Diniz2011}, quantitative predictions
for the system dynamics require to know not only the overall linewidth
but also the detailed shape of $\rho(\omega)$. In particular Gaussian
and Lorentzian distributions yield very different results. Spin ensembles
inhomogeneously broadened by dipolar interactions are expected to
show a Lorentzian lineshape \cite{dobrovitski2008} with a cutoff;
however complications in our experiment arise due to hyperfine coupling
with the $^{14}N$ atom nuclear spin, and to a possible misalignment
of $B_{\mathrm{NV}}$ with the $[1,0,0]$ crystalline axis causing
the four NV orientations to have slightly different frequencies. As
a result, we assume a phenomenological lineshape for our spin ensemble
adjusted for reaching good agreement both with spectral and time-domain
experimental data, as a convolution of a Gaussian and a Lorentzian
profile with respective widths $\sigma$ and $\gamma$. The parameters
chosen in the following ($\sigma/2\pi=5.12$~MHz and $\gamma/2\pi=1$~MHz)
yield a lineshape $\rho(\omega)$ compatible with ODMR data although
slightly broader, as can be seen in Fig.\ref{Fig2}e (see Supplementary
Material). Using such distribution and the formulas above, we obtain
quantitative agreement for spectroscopic (see Fig. \ref{Fig1}e) as
well as time-domain (see Fig. \ref{Fig2}d) measurements.

We also study the dependence of the microwave field exchange on the
resonator-spins detuning by measuring $a(\tau)$ for various $\Delta\Phi$
(see Fig.\ref{Fig3}). Out of resonance, the oscillations amplitude
is reduced and their frequency increases as expected for two coupled
oscillators. The asymmetry in the data between the $\Delta\Phi<\overline{\Delta\Phi}$
and $\Delta\Phi>\overline{\Delta\Phi}$ sides is an artefact of the
nonlinear dependence of $\omega_{\mathrm{r}}$ on $\Delta\Phi$ and
of a residual hybridization of the resonator with the spins caused
by the finite initial detuning $\left[\omega_{r}(\Phi_{OFF})-\omega_{\mathrm{-}}\right]/g\sim3$,
and is well reproduced by theory. The largest discrepancy is observed
for pulse amplitudes $\Delta\Phi\sim90m\Phi_{0}$, where additional
features are clearly seen in the experiment. We attribute them to
a small density of NV centers having a $^{13}C$ among their closest
neighbor, known to shift the electron spin frequency by $\pm65$~MHz
due to the hyperfine interaction, and that was not taken into account
in the calculation. 

\begin{figure}
\includegraphics[width=7cm]{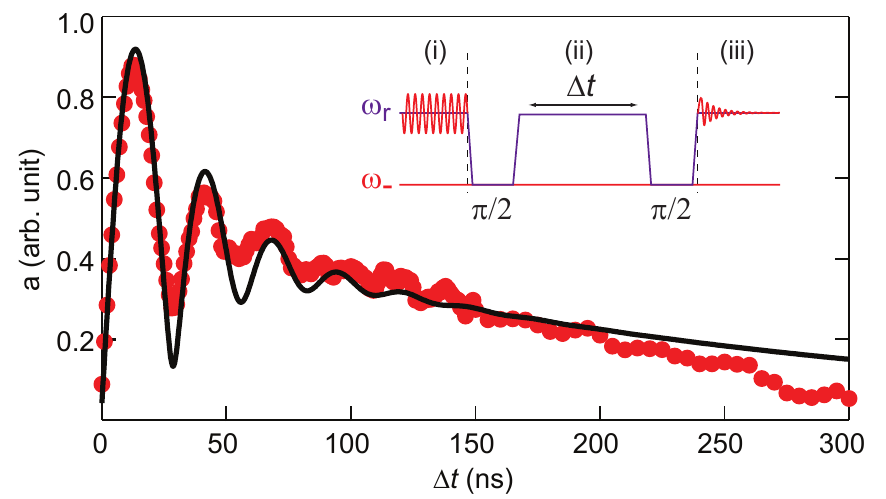}\caption{Ramsey fringes experiment. The normalized microwave amplitude $a(\Delta t)$
after the pulse sequence shown in inset is plotted. Red dots are experimental
data, black continuous line is theory. }

\label{Fig4}%
\end{figure}

We finally perform a Ramsey-like experiment in order to quantify the
time during which the microwave field can be stored in the spin ensemble.
Starting as before with a steady-state microwave field in the resonator
biased at $\Phi_{\mathrm{OFF}}$, the resonator is tuned in resonance
with the spins at $t=0$ for a $\pi/2$ pulse ($g_{\mathrm{ens}}\tau=\pi/2$),
after which it is detuned by $\Delta\omega/2\pi=30$~MHz during $\Delta t$,
then tuned back to resonance for a second $\pi/2$ pulse, and finally
tuned back at $\omega_{\mathrm{r}}(\Phi_{\mathrm{OFF}})$, after which
$a(\Delta t)=\overline{A(2\tau+\Delta t+t_{\mathrm{off}})}$ is measured.
Due to the beating between the effective spin oscillator and the resonator,
oscillations are observed in $a(\Delta t)$ at frequency $\Delta\omega$
as seen in Fig.\ref{Fig4}. The oscillations amplitude decay time
$\sim30$~ns gives the storage time, on the order of $\sim2/\Delta$
as expected. The full calculation, performed as explained above, yields
again reasonable agreement with the measurements. 

As shown by the previous results, inhomogeneous broadening appears
as a serious obstacle to the successful implementation of hybrid quantum
circuits based on spin ensembles. One first obvious solution is to
obtain samples with narrower inhomogeneous linewidths. In the specific
case of NV centers, this requires a very efficient conversion rate
$\eta$ of P1 centers initially present in the diamond crystal in
NV centers. Improvement by one order of magnitude compared to the
sample used in this work (where $\eta$ is a few percent) is within
reach. This would reduce the inhomogeneous linewidth by the same amount,
at which point inhomogeneous broadening would be dominated by the
NV centers hyperfine structure ($\Delta/2\pi\sim4$~MHz for $^{14}N$).
Another solution is to increase the NV concentration even further
to reach the regime $g_{\mathrm{ens}}\gg\Delta$ and rely on the cavity
protection effect. Finally, it should be possible, by applying refocusing
techniques used for atomic ensemble-based quantum memories \cite{Lvovsky2009},
to recover the quantum information lost in the dark states into the
superradiant state and then into the resonator. This would be the
best way to fully benefit from the spin superior coherence properties. 

In conclusion we have demonstrated the coherent storage and retrieval
of a classical microwave field of $\sim100$ photons from a superconducting
resonator into collective excitations of an ensemble of NV centers
and discussed the implications of inhomogeneous broadening. This work
opens the way to the operation of a quantum memory in which a single
microwave photon would be coherently stored and retrieved from a superconducting
qubit into the spin ensemble.

We gratefully thank P. Sénat, J.-C. Tack, P.-F. Orfila, M. de Combarieu,
P. Forget, and P. Pari for priceless technical help, and we acknowledge
useful discussions within the Quantronics group. We acknowledge the
support of European Contracts MIDAS and SOLID.

\part*{Supplementary Material}

\section{Measurement Setup}

\begin{figure*}[t]
\includegraphics[width=18cm]{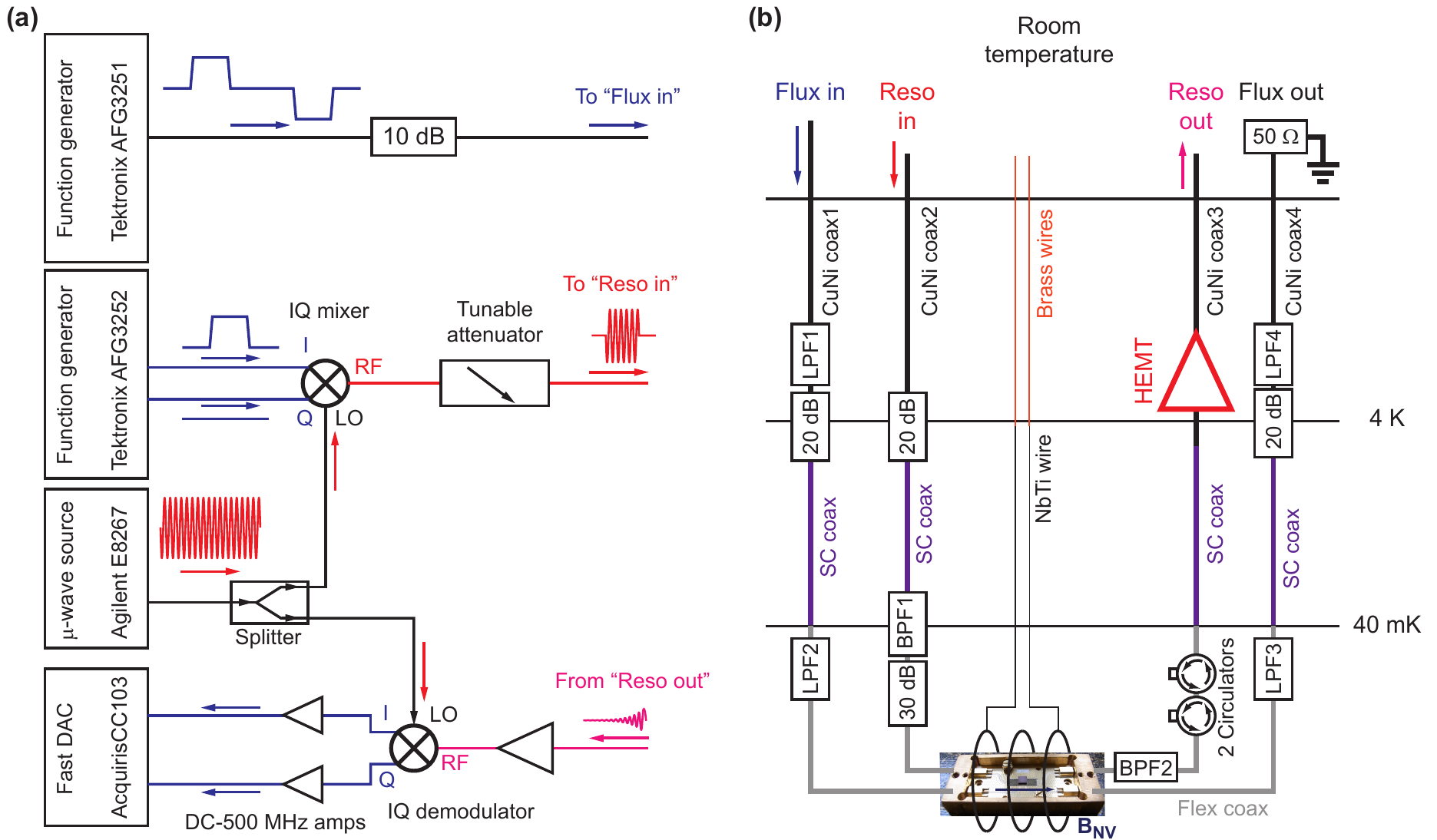}

\caption{Measurement setup and wiring. (a) Full configuration of the measurement
apparatus at room temperature. (b) Schematic of the wiring inside
the dilusion refrigerator. LPF1, LPF2, LPF3, and LPF4 are low-pass
filters having cutoff frequencies 1.35 GHz, 450 MHz, 450 MHz, and
1.35 GHz, respectively. BPF1 actually consists of two series of low-pass
(5.4 GHz) and high-pass (1.3 GHz) filters, and BPF2 is a band-pass
filter (2.5 to 4 GHz). CuNi coax1, 2, and 4 are coaxial cables made
of CuNi, and CuNi coax3 is a silver-plated CuNi coaxial cable. SC
coax is a superconducting NbTi coaxial cable. Flex coax is a low-loss
flexible coaxial cable. The two circulators are STE-1502KS (Pamtech).
Here the small rectangles represent ports terminated by 50 $\Omega$.
HEMT is a cryogenic microwave amplifier having low-noise temprature
(Caltech). A DC current for the DC magnetic field $B{}_{NV}$ is applied
through an RC low-pass filter at room temperature.}

\label{FigSetup}%
\end{figure*}

Figure \ref{FigSetup}(a) shows the full configuration of the room
temperature apparatus, and \ref{FigSetup}(b) the wiring inside the
fridge. The fast current pulse is generated by an arbitrary function
generator AFG3251 (Tektronix) and input into the fridge from the port
{}``Flux in'' through a 10 dB attenuator. After being filtered by
low pass filters (LPF1 and LPF2) at each temperature stage and further
attenuated by 20 dB at 4 K stage, the flux pulse passes through the
flux line {[}see the solid arrows in Fig. \ref{FigPulse}(a) and (b){]},
and goes back to room temperature through low-pass filters (LPF3 and
LPF4) and a 20 dB attenuator. The end of the fast flux circuit ({}``Flux
out'') is terminated by a broadband 50 $\Omega$. 

The microwave pulse is generated by mixing continuous microwave and
a fast DC voltage pulse using an IQ-mixer\cite{IQ}. The microwave
pulse is input to the port {}``Reso in'' after being attenuated
by a tunable attenuator, and reaches to the input port of the chip
through attenuators at each temperature stage and a band-pass pass
filter (BPF1) at 40 mK. After being filtered by a band pass filter
(BPF2) and isolated by two circulators, the transmitted signal is
amplified by a cryogenic HEMT (high electron mobility transistor)
amplifier having noise temperature $\sim$ 4 K. The amplified signal
is then further amplified at room temperature by two microwave amplifiers,
and demodulated into in-(I) and quadrature-(Q) phase components using
an IQ-demodulator. The demodulated signal is sampled at 500 MS/s by
a fast data acquisition card Acqiris CC103 (Agilent) after being further
amplified by two amplifiers. 

The resonator chip fixed onto a printed circuit board is mounted inside
a sample box made of copper {[}Fig. \ref{FigSetup}(b){]}. A DC magnetic
field $B{}_{NV}$ is applied parallel to the chip by passing a DC
current through an outer superconducting coil. The sample box and
the coil are surrounded by two magnetic shieldings consisting of permalloy
tape VC6025X (VacuumSchmelze) and a lead cylinder {[}not shown in
Fig. \ref{FigSetup}(b){]}. 

\section{Flux pulse}

\begin{figure}[h]
\includegraphics[width=8cm]{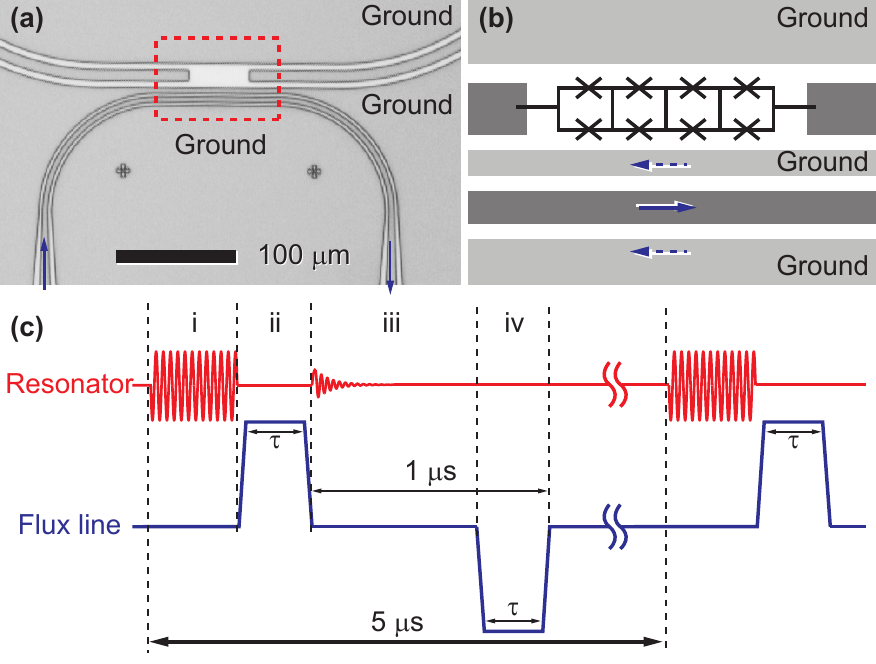}\caption{(a) Photograph of the area where the 4-SQUIDs array is placed in the
resonator. Note that this photograph was taken before the SQUIDs were
fabricated. The arrows shows a flow of an applied fast current pulse.
(b) Schematic of inside the dashed box in (a) (here the SQUIDs are
drawn). The applied (return) current pulses is represented by a solid
(dashed) arrow(s). (c) Actual pulse sequence for the measurement:
(i) excitation, (ii) interaction, (iii) data accumulation, and (iv)
compensation in the sequence. After data sampling (iii), an inverted
pulse (iv) is applied to compensate the time integral to $0$ (see
text). The whole length of one sequence is 5 $\mu$s.}

\label{FigPulse}%
\end{figure}

Here we describe a few technical details on the flux pulse used to
tune the resonator frequency at the nanosecond scale. As shown in
Fig. \ref{FigPulse}, the pulse is applied using an arbitrary function
generator AFG3252 (Tektronix) that sends a current pulse through an
on-chip line. The risetime of the voltage pulse is $2$~ns. Passing
current through the on-chip antenna generates screening currents through
the ground planes, as shown by the dashed arrows in Fig. \ref{FigPulse}(b),
of our superconducting circuit with very low damping times (of the
order of the ms) that in turn affect the flux through the SQUID. As
a result, the SQUID offset flux bias point was found to depend on
the time integral of the flux pulse. In addition, the flux applied
to the SQUID loop by sending a fast current pulse through the flux
line was strongly reduced (by a factor $\sim50$) compared to the
flux applied with a dc current of the same amount. We measure this
current-to-flux transfer function with calibration experiments, allowing
us to convert a voltage pulse into flux as in figure $3$ of the main
text. In order for the bias point to be independent of the amplitude
and duration of the pulse, we had to add, at the end of each sequence,
a compensation flux pulse opposite to the first one {[}see Fig. \ref{FigPulse}(c){]}.
Such a compensation pulse has strictly no incidence on the experiment
outcome since it is applied long after the microwave signal is detected. 

\section{Theory details}

We now give more details on the theory calculations. All these calculations
are performed in the Holstein-Primakoff approximation, in which the
spins and the resonator are described by harmonic oscillators, as
explained in the text. The system Hamiltonian is $H/\hbar=\omega_{\mathrm{r}}(\Phi)a^{\dagger}a+\sum\omega_{\mathrm{j}}b_{\mathrm{j}}^{\dagger}b_{\mathrm{j}}+\sum ig_{\mathrm{j}}(b_{\mathrm{j}}^{\dagger}a-b_{\mathrm{j}}a^{\dagger})$,
$g_{\mathrm{j}}$ being the coupling constant of spin $j$ with the
resonator.

\subsection{Rabi oscillations}

We first calculate the oscillating signal resulting from the cycles
of storage and retrieval of the resonator field of amplitude $\alpha$
into the spin ensemble (fig. 2d in the article). A first remark is
that the system consisting of coupled harmonic oscillators, its time
evolution does not depend on the initial field amplitude. As a result
we only calculate $\alpha_{Rabi}=\left\langle 0\right|a(t)a^{\dagger}(0)\left|0\right\rangle $
which represents the probability amplitude that a photon created at
$t=0$ is still present at time $t$. As shown in \cite{Diniz2011-1}
this quantity can be calculated by considering an effective non-Hermitian
Hamiltonian 

\begin{equation} H_{eff}/\hbar = \left( \begin{array}{cccc} \tilde\omega_0 & i g_1 & i g_2 & \ldots \\ -i g_1 & \tilde\omega_1 & & \\ -i g_2 & & \tilde\omega_2 &\\ \vdots & & & \ddots \\ \end{array} \right)\, . \end{equation}

with complex angular frequencies $\tilde{\omega}_{0}=\omega_{r}-i\kappa/2$
and $\tilde{\omega}_{k}=\omega_{k}-i\gamma_{0}/2$ ; here, $\gamma_{0}$
is the spontaneous emission rate of each spin (that we take here to
be zero since NV centers at low temperature have negligible energy
relaxation). Indeed, introducing the vector $X(t)$ of coordinates
$\left[\left\langle a(t)a^{\dagger}(0)\right\rangle ,...,\left\langle b_{j}(t)a^{\dagger}(0)\right\rangle ,...\right]$
it can be shown that $dX/dt=-(i/\hbar)H_{eff}X$. The formal solution
to this equation is then 

\begin{equation} X(t) = \mathcal{L}^{-1}  [ (s+i H_{eff} /\hbar)^{-1}  X(0)] \, , \end{equation}

with $X(0)=x_{G}$ and $x_{G}\equiv(1,0,...,0)$ . This implies that
$\alpha(t)=x_{G}{}^{\dagger}\cdot X(t)=\mathcal{L}^{-1}\left[t_{1}(s)\right]$
with $t_{1}(s)=x_{G}{}^{\dagger}\cdot(s+iH_{eff})^{-1}\cdot x_{G}$
and $\mathcal{L}[f(s)]=\int e^{-st}f(t)dt$, $s$ being a complex
number. Since $t_{1}(s)$ is not singular on its imaginary axis, we
only need $t_{1}$ for pure imaginary argument $s=-i\omega$ to perfom
the Laplace transform inversion. As shown in \cite{Diniz2011-1},
we have $t_{1}(-i\omega)=i/\left[\omega-\omega_{0}+i\kappa/2-W(\omega)\right]$
with $W(\omega)=g_{ens}^{2}\int\rho(\omega')d\omega'/\left[\omega-\omega'+i\gamma_{0}/2\right]$.
Computing $\alpha(t)$ is thus achieved by evaluating $t_{1}$ for
the distribution $\rho(\omega)$ described in the main text, and numerically
evaluting the inverse Laplace transform. This calculation also enables
us to calculate the transmission as plotted in Fig.1e, because as
shown in \cite{Diniz2011-1}, the resonator transmission is obtained
as $t(\omega)=-\left(\kappa/2\right)t_{1}(-i\omega)$.

An additional complication in the experiment is that the resonator-spin
detuning $\delta$ during the microwave pulse is finite, implying
that the initial state is actually already hybridized. To take this
into account in the calculation, we make the approximation that this
initial state is a coherent superposition of $x_{G}$ and $x_{S}\equiv(0,g_{1},...,g_{j},..,g_{N})/g_{ens}$.
The vector $x_{s}$ is associated with the excitation of the superradiant
mode. Thus the initial state is written $x(t=0)=\cos(\theta/2)x_{G}+i\sin(\theta/2)x_{S}$
with mixing angle $\tan\theta=2g_{ens}/\delta$. Using a similar analysis,
we now have $\alpha_{Rabi}=\cos(\theta/2)\mathcal{L}^{-1}\left[t_{1}(s)\right]+i\sin(\theta/2)\mathcal{L}^{-1}\left[t_{4}(s)\right]$.
In addition to $t_{1}(s)$ we then need $t_{4}(s)=x_{G}^{\dagger}(s+iH_{eff})^{-1}x_{S}$,
as shown in \cite{Diniz2011-1}, $t_{4}(-i\omega)=-it_{1}(-i\omega)W(\omega)/g_{ens}$.

\subsection{Ramsey fringes}

Each $\pi/2$ pulse necessary to realize the Ramsey-like experiment
is realised by bringing the resonator and spins to resonance. For
a fast pulse, the resonant interaction maps continuously the coherent
state of the field to the superradiant mode in the spins. We calibrate
the interaction time in such a way as to transform the state $x_{G}$
into the superposition $\frac{x_{G}-x_{S}}{\sqrt{2}}$. After the
first $\pi/2$ pulse, the resonator is kept detuned from the spin
ensemble for a time $t$. The system state at this point can be evaluated
using eq. (2). We define $X_{G}(t)$ (resp. $X_{S}(t)$) as the vector
of coordinates $\left[\left\langle a(t)a^{\dagger}(0)\right\rangle ,...,\left\langle b_{j}(t)a^{\dagger}(0)\right\rangle ,...\right]$
at time $t$ with initial conditions $x_{G}$ (resp. $x_{S}$). A
second $\pi/2$ pulse is then applied before the amplitude $\alpha_{RF}$
of the field in the resonator is measured :

\begin{equation}\begin{split}    \alpha_{RF} &= \frac{1}{\sqrt 2} x_G^\dagger \cdot U_{\pi/2} (X_G(t) - X_S(t) )  \\ 
  &= \frac{1}{2} (x_G^\dagger + x_S^\dagger ) \cdot (X_G(t) - X_S(t) ) \\%
  &=\frac{1}{2} \mathcal{L}^{-1} ( t_1(s) - t_2(s) + t_3(s) - t_4(s) ) \, ,\end{split}\end{equation} where $t_{2}(s)=x_{S}{}^{\dagger}\cdot(s+iH_{eff})^{-1}\cdot x_{S}$
and $t_{2}(s)=x_{S}{}^{\dagger}\cdot(s+iH_{eff})^{-1}\cdot x_{G}$,
$t_{1}$ and $t_{4}$ are defined above. As shown in \cite{Diniz2011-1},
$t_{2}(-i\omega)=-t_{1}(-i\omega)W(\omega)(s+i\tilde{\omega}_{0})/g_{ens}^{2}$
and $t_{3}=-t_{4}$.

\subsection{Density distribution}

The spins density distribution that we use is the convolution of a
normalized Gaussian of standard deviation $\sigma$ with a normalized
Lorentzian of HWHM $\gamma$. The resulting distribution is known
as the Voigt profile and is also normalized.

\begin{equation}\begin{split} \rho(\omega'; \sigma , \gamma) &= \int_{-\infty}^\infty G(\omega'';\sigma) L(\omega' - \omega'' ;\gamma) d\omega' \\%
&= \frac{1}{\sigma \sqrt{2 \pi}}  \frac{\gamma}{ \pi} \int_{-\infty}^\infty \frac{e^{\omega''^2/2 \sigma^2} }{ (\omega'- \omega'')^2 + \gamma^2} \; d\omega' \, . \end{split}\end{equation} 

An important property is that we can compute analytically the function
$W(\omega)$ :

 \begin{equation}\begin{split} W(\omega; \sigma, \gamma , \gamma_0 ) = \frac{\Omega^2}{\sigma \sqrt{2 \pi}}  \frac{\gamma}{ \pi} \int \int \frac{e^{\omega''^2/2 \sigma^2} }{ (\omega'- \omega'')^2 + \gamma^2} \frac{d\omega' d\omega''}{ \omega - \omega'+ i \gamma_0} \; ,\end{split}\end{equation}.

Taking the limit $\gamma_{0}\rightarrow0$ we have

 \begin{equation}\begin{split}   W(\omega; \sigma, \gamma) = -i \sqrt\frac{\pi}{2} \; \frac{ \Omega^2 }{\sigma } \;   \exp \left\{-\left( \frac{\omega + i\gamma}{ \sigma \sqrt{2}} \right)^2 \right\} \text{erfc} \left( - i \frac{\omega + i\gamma}{ \sigma \sqrt{2}} \right) \; ,\end{split}\end{equation}.

The function that is taken in the theory plots of the main text actually
takes into account the well-known NV center hyperfine splitting due
to $^{14}N$ by adding up three identical distributions $\rho(\omega)$
separated by $2.2$~MHz \cite{Acosta2009}. The resulting distribution
is the one shown in figure $2$ of the main text. This is conviently
done since $W[\rho_{1}+\rho_{2}]=W[\rho_{1}]+W[\rho_{2}]$ for any
two different spin distributions $\rho_{1,2}$. Using this formula
for $W(\omega)$ we can evaluate the functions $t_{i}$ necessary
for the simulation of the Rabi and Ramsey-like experiments.


\begin{thebibliography}{24}
\bibitem{IQ} Here a DC constant voltage is applied to the Q-port
of the IQ-mixer for the perpose of cancelling the offset, i.e., to
prevent the microwave signal from leaking out during the pulse is
off.
\end{thebibliography}
\end{document}